\documentclass[aps,prc,twocolumn,showpacs,amsmath,amssymb,nofootinbib,10pt]{revtex4-2}
\usepackage{graphicx}
\usepackage{amsfonts, amsmath, amssymb}
\usepackage{color}
\usepackage{times}
\usepackage[utf8]{inputenc}
\usepackage{bm}
\usepackage[normalem]{ulem}
\usepackage{multirow}
\usepackage{url}
\usepackage{natbib}
\usepackage{mathrsfs}
\usepackage{physics}
\usepackage{comment}
\usepackage[nointegrals]{wasysym}
\usepackage{booktabs}
\usepackage[table,xcdraw]{xcolor}
\usepackage[caption=false]{subfig}
\usepackage[colorlinks=true,citecolor=blue,urlcolor=blue,linkcolor=blue]{hyperref} 
\begin{document}
\title{Bayesian Inference of Fermionic Dark Matter Admixed Neutron Stars Informed by Galaxy Cluster Constraints}
\author{H. C. Das$^1$}
\author{J. B. Wei$^2$}
\affiliation{$^1$Institut f\"ur Theoretische Physik, Goethe Universit\"at, Max-von-Laue-Str.~1, D-60438 Frankfurt am Main, Germany}
\affiliation{$^2$Physics Department, University of Geosciences, Wuhan, P.R.~China}
\newcommand{\HCD}[1]{\textsf{\color{blue}{#1}}}
\newcommand{\JBW}[1]{\textsf{\color{orange}{#1}}}
\newcommand{\JSB}[1]{\textsf{\color{red}{#1}}}
\date{\today}
\begin{abstract}
Self-interacting dark matter is constrained on galactic and cluster scales by its elastic scattering cross section per unit mass, $\sigma/m_\chi \sim 0.1-10\ {\rm cm^2/g}$. We incorporate this cross section into the equation of state of an asymmetric, self-interacting fermionic dark matter component and perform a Bayesian inference of dark matter admixed neutron stars in the two-fluid framework at $\sigma/m_\chi = 0.1$, $1$ and $10\ {\rm cm^2/g}$. Seven nuclear matter parameters are sampled together with the dark particle mass and mass fraction, and confronted with nuclear theory and neutron star observations. The nuclear sector and the global observables are statistically indistinguishable across the whole range of cross sections, while the data constrain the dark matter mass to $m_\chi \simeq 1.45-1.88$~GeV and its mass fraction to $f_\chi \simeq 0.05-0.09$, depending on the cross section. Finally, comparing runs built from the same parameter draws, so that only the cross section differs, we find that it does change the baryonic radius appreciably, but only for light dark matter, a region the astrophysical data largely disfavor. The self-interaction is therefore not too weak to be seen; however its effect is hidden in the region of parameter space that the present observational data already disfavor.
\end{abstract}
\maketitle
\section{Introduction}
\label{sec:intro}
The existence of dark matter (DM) is established by a broad range of evidence, such as galactic rotation curves~\cite{Rubin:1970.ApJ.159.379}, the dynamics of galaxy clusters~\cite{Zwicky:2009.GRG.41.207,Trimble:1987.ARAA.25.425}, gravitational lensing~{\cite{Zwicky:1937.PR.51.290,Clowe:astro-ph.0608407}, and the cosmic microwave background~\cite{Planck:1807.06209,Bergstrom:hep-ph.0002126,Bertone:hep-ph.0404175,Feng:1003.0904,Bertone:1605.04909}. Its particle nature, however, remains unknown, and the null results of direct detection experiments~\cite{LZ:2207.03764,XENON:2303.14729} have motivated the search for astrophysical environments in which DM may accumulate and reveal itself indirectly. Compact stars are among the most promising of these environments~\cite{Goldman:1989.PRD.40.3221, Kouvaris:0708.2362, Bertone:0709.1485, DeLavallaz:1004.0629, Kouvaris:1111.4364, Kouvaris:1507.00959, Bramante:2024.PhysRep.1052.1, Grippa:2412.09381}, because their extreme density and gravity enhance the capture of DM particles. The resulting admixed configurations may have masses, radii, and tidal deformabilities that differ measurably from those of purely baryonic stars~\cite{Sandin:0809.2942,Ciarcelluti:1005.0857,Leung:1205.1909,Nelson:1803.03266,Ivanytskyi:2109.03801,Das:2002.00594,Das:2109.01853,Das:2112.14198,Liu:2403.17024,Pitz:2025.PRD.111.043050}.

A particularly well-motivated scenario is asymmetric dark matter (ADM), whose relic abundance is set by a particle--antiparticle asymmetry rather than by thermal freeze-out, so that there is no annihilation inside the star~\cite{Zurek:1308.0338,Kouvaris:1507.00959,Ellis:1804.01418}. If, in addition, the dark sector has a repulsive self-interaction mediated by a vector boson, the resulting equation of state (EOS) becomes considerably stiffer than that of a free Fermi gas~\cite{Narain:astro-ph.0605724}, and the DM component can support a significant amount of mass on its own. Such self-interacting dark matter (SIDM)~\cite{Spergel:astro-ph.9909386,Tulin:1705.02358} is independently motivated on astrophysical grounds: it alleviates the core--cusp and too-big-to-fail problems of collisionless cold DM on dwarf-galaxy scales~\cite{Loeb:1011.6374,Oh:1011.0899,BoylanKolchin:1103.0007,Bullock:1707.04256}, while remaining compatible with the constraints derived from colliding galaxy clusters~\cite{Markevitch:astro-ph.0309303,Randall:0704.0261,Harvey:1503.07675}.

These astrophysical systems, however, do not constrain the coupling and the mediator mass separately, but only the elastic scattering cross section per unit DM mass. Observations of the Bullet Cluster, of merging galaxy groups and clusters, and of dwarf-galaxy cores bracket this quantity in the range~\cite{Markevitch:astro-ph.0309303,Kaplinghat:1508.03339,Sagunski:2006.12515,Loeb:2203.11962,Tulin:1705.02358}
\begin{equation}
\sigma/m_\chi \sim 0.1-10\ {\rm cm^2/g}\,,
\end{equation}
with the larger values favored at the lower velocities characteristic of dwarf galaxies. This is an observational input of a kind quite different from the neutron star data usually employed in Bayesian inference.

Bayesian inferences of dark matter admixed neutron stars (DANSs) within the two-fluid framework have been performed by several groups, using neutron-star observables such as masses, radii, and tidal deformabilities to constrain the dark-matter particle mass and its mass fraction within the star~\cite{Ivanytskyi:2208.03282, Rutherford:2410.00140, Arvikar:2506.20736, Liu:2506.08376, Arvikar:2512.13574, Rutherford:2609.14494, Karkevandi:2609.38055}. More recently, the properties of DANSs have also been investigated either in an agnostic framework~\cite{karan:2607.03840} or through parameter-space studies incorporating multimessenger observational constraints~\cite{Liu:2307.11313, Thakur:2024.PRD.109, Liu:2403.17024, Mariani:2311.14004, Kumar:2501.16829, Rather:2609.03964}, for both fermionic and bosonic dark-matter models. In these studies, the dark-sector parameters are constrained primarily by neutron-star observables. Although Refs.~\cite{Mariani:2311.14004, Kumar:2501.16829} additionally restrict the dark-sector parameter space using self-interaction cross sections inferred from the dynamics of the Bullet Cluster, Abell clusters, and dwarf galaxies, these studies do not perform a Bayesian analysis. Conversely, the fully Bayesian analysis of fermionic dark matter by Rutherford {\it et al.}~\cite{Rutherford:2410.00140} does not incorporate external constraints on the dark sector. A Bayesian framework that simultaneously incorporates neutron-star observables and independent astrophysical constraints on the dark sector therefore remains to be explored.

The purpose of the present study is to treat the self-interaction cross section, $\sigma/m_\chi$, as an additional observational input that enters directly into the DM EOS. This allows us to address two complementary questions; ``can neutron-star observables provide information on the DM self-interaction inferred from colliding galaxy clusters, and what observable signatures of DM self-interactions might be imprinted on neutron stars?" This approach is possible within the Born approximation, in which the self-interaction cross section can be expressed in terms of the dark-sector coupling and mediator mass~\cite{Tulin:1302.3898, Zurek:1308.0338, Kouvaris:1507.00959, Maselli:1704.07286}. Once $\sigma/m_\chi$ is fixed, the DM EOS depends only on the DM particle mass. We exploit this property to perform three complete Bayesian inferences, for $\sigma/m_\chi = 0.1$, $1$, and $10~{\rm cm^2/g}$, thereby spanning the astrophysically relevant range and enabling a consistent comparison of the resulting constraints under otherwise identical assumptions.

The paper is organized as follows. Section~\ref{sec:form} presents the normal matter and DM EOSs, the two-fluid stellar structure equations, and the Bayesian setup, including the prior-predictive run. Section~\ref{sec:results} presents the inferred nuclear and dark sector parameters, the global observables, the core/halo structure of the admixed configurations, an explicit analysis of where the sensitivity to the dark sector resides, the evidence for or against dark halos, the region of parameter space in which the self-interaction would be visible, and the radial stability of the configurations. We summarize our conclusions in Section~\ref{sec:conclusion}.
\section{Formalism}
\label{sec:form}
\subsection{Normal Matter EOS}
\label{subsec:NM_EOS}
For the normal matter (NM) EOS, we use the relativistic mean-field (RMF) model, in which the nuclear interaction is mediated by the exchange of the $\sigma$, $\omega$, and $\rho$ mesons, including their self- and cross-couplings. The energy density of the system is~\cite{Kumar:2005.08320, Pradhan:2209.12657}
\begin{align}
\varepsilon_{\rm nucl.} & = \sum_{\rm N = p,n} \frac{1}{\pi^2}\int_{0}^{k_N} k^2 \sqrt{k^2+m_N^{*2}} \, dk
\nonumber\\
&
+m_{\sigma}^2{\sigma}^2\Bigg(\frac{1}{2}+\frac{\kappa_{3}}{3!}\frac{g_\sigma\sigma}{m_N}+\frac{\kappa_4}{4!}\frac{g_\sigma^2\sigma^2}{m_N^2}\Bigg)+\frac{1}{2}m_{\omega}^2\,\omega^2
\nonumber\\
&
+ \frac{1}{2}m_{\rho}^2\,\rho^{2}
+ \frac{1}{8}\zeta_{0}\,{g_{\omega}^2}\,\omega^4
+ 3\Lambda_{\omega}\, g_\omega^2\, g_\rho^2\, \omega^2 \rho^2\,,
\label{eq:eden_rmf}
\end{align}
where $k_N$ and $m_N^* = m_N - g_\sigma\sigma$ are the Fermi momentum and the effective mass of the nucleons, respectively. $g_\sigma$ ($m_\sigma$), $g_\omega$ ($m_\omega$), and $g_\rho$ ($m_\rho$) are the coupling constants (masses) of the $\sigma$, $\omega$, and $\rho$ mesons, respectively. The nucleon mass $m_N$ is taken as 939~MeV; $\kappa_3$ and $\kappa_4$ are the self-coupling constants of the $\sigma$ meson and $\zeta_0$ that of the $\omega$ meson; $\Lambda_\omega$ is the $\omega$--$\rho$ cross-coupling. The masses of the $\sigma$, $\omega$, and $\rho$ mesons are taken as 500, 783, and 770~MeV, respectively.

In addition to nucleons, NSs contain a small fraction of leptons, which are needed for charge neutrality in $\beta$-equilibrium. Their energy density is
\begin{equation}
\varepsilon_{\rm lept.} = \sum_{\rm l = e, \mu} \frac{1}{\pi^2}\int_0^{k_l} k^2 \ \sqrt{k^2+m_l^2} \ dk \, ,
\label{eq:eden_l}
\end{equation}
where $m_l$ is the lepton mass ($m_e = 0.511$~MeV and $m_\mu = 105.66$~MeV). The NM energy density is the sum of the nucleon and lepton contributions, $\varepsilon_{\rm N} = \varepsilon_{\rm nucl.} + \varepsilon_{\rm lept.}$, and the pressure follows from the thermodynamic relation $p_{\rm N} = \sum_i\mu_i n_i - \varepsilon_{\rm N}$, where $\mu_i$ and $n_i$ are the chemical potentials and number densities of the nucleons and leptons.

To obtain the RMF EOS, the values of the coupling constants are required. In this study, we adopt the approach in which the coupling constants are calculated directly from the nuclear matter properties (NMPs), as discussed in detail in Refs.~\cite{Chen:1408.4159, Hornick:1808.06808, Zhu:2305.16058}. For a fixed model, the unknown parameters in Eq.~(\ref{eq:eden_rmf}) are $g_\sigma$, $g_\omega$, $g_\rho$, $\kappa_3$, $\kappa_4$, $\zeta_0$, and $\Lambda_\omega$. These are obtained from the NMPs, namely the binding energy ($E_{\rm sat}$), incompressibility ($K_{\rm sat}$), symmetry energy ($J_{\rm sat}$), its slope ($L_{\rm sat}$), and the nucleon effective mass ($M^*_{\rm sat}$) at the saturation density ($n_{\rm sat}$). We calculate all coupling constants at saturation density following the approach of Ref.~\cite{Chen:1408.4159}, except $\zeta_0$, which has a significant impact on the high-density behavior of the nuclear matter probed inside a NS. The value of $\zeta_0$ controls the softness or stiffness of the EOS at supra-saturation densities and hence directly affects the NS maximum mass. It is therefore treated as a free parameter with its own prior. The remaining parameters of the model are $n_{\rm sat}$, $E_{\rm sat}$, $K_{\rm sat}$, $J_{\rm sat}$, $L_{\rm sat}$, and $M^*_{\rm sat}$. Their prior ranges are given in Table~\ref{tab:prior_table}.
\subsection{Dark Matter EOS}
\label{subsec:DM_EOS}
In this study, the asymmetric dark matter (ADM) is described by self-interacting fermions that neither annihilate nor interact with ordinary baryonic matter. The self-interaction between DM particles is mediated by a vector boson of mass $m_\phi$ with coupling strength $g_\chi$. The energy density and pressure of this model are then given by~\cite{Narain:astro-ph.0605724,Liu:2403.17024,Barbat:2404.12875,Das:2509.04831}
\begin{align}
\varepsilon_{\rm D} &=
\frac{m_\chi^4}{8\pi^2}
\Big[ x\sqrt{1+x^2}\left(2x^2+1\right) - \mathop{\text{arcsinh}}(x) \Big]
+ \delta \:,
\\
p_{\rm D} &= \frac{m_\chi^4}{8\pi^2}
\Big[ x\sqrt{1+x^2}\left(\frac{2x^2}{3}-1\right) + \mathop{\text{arcsinh}}(x) \Big]
+ \delta \:,
\label{eq:EOS_DM}
\end{align}
where
\begin{equation}
x = \frac{k_\chi}{m_\chi} = \frac{\left(3\pi^2 n_\chi\right)^{1/3}}{m_\chi}
\end{equation}
is a dimensionless kinetic parameter expressed in terms of the DM Fermi momentum $k_\chi$ and number density $n_\chi$. Here $\chi$ denotes the fermionic DM particle, with a mass $m_\chi$ ranging from the MeV to the GeV scale, as discussed in Refs.~\cite{Kain:2102.08257,Nelson:1803.03266,Miao:2204.05560,Liu:2403.17024}. The self-interaction term reads
\begin{equation}
\delta = \left( \frac{y\, n_\chi}{m_\chi} \right)^2\, , \quad y \equiv \frac{g_\chi m_\chi}{(\sqrt{2}\,m_\phi)}  \:.
\label{eq:y}
\end{equation}
The scaling relations governing the EOS and the mass-radius relations of pure fermionic DM stars in different interaction regimes have been discussed in Ref.~\cite{Narain:astro-ph.0605724}.

The parameters $m_\chi$ and $y$ are not independent; they are jointly constrained by the observational bounds on the DM self-interaction cross section $\sigma$ derived from studies of colliding galaxy clusters~\cite{Markevitch:astro-ph.0309303,Kaplinghat:1508.03339,Sagunski:2006.12515,Loeb:2203.11962},
\begin{equation}
\sigma/m_\chi \sim (0.1-10) \frac{\rm cm^2}{\rm g} \approx (0.1-10) \times 4560\;\text{GeV}^{-3}\:.
\end{equation}
It has been shown~\cite{Tulin:1302.3898,Zurek:1308.0338,Kouvaris:1507.00959,Maselli:1704.07286} that the Born approximation
\begin{equation}
\frac{\sigma}{m_\chi} = \frac{y^4}{\pi\, m_\chi^3}
\end{equation}
remains valid in the limit $y \rightarrow 0$ for larger masses. This constraint implies
\begin{align}
y = \left(\pi \times 4560 \times \frac{\sigma}{m_\chi}\right)^{1/4} \overline{m}_{\chi}^{3/4},
\label{eq:ymu}
\end{align}
where $\overline{m}_{\chi} \equiv m_\chi / \text{GeV}$. With this relation imposed, the DM EOS depends only on $m_\chi$ once $\sigma/m_\chi$ has been fixed.

It is useful to show how the cross section enters the stellar structure problem, since it controls the magnitude of every effect reported below. Substituting Eq.~(\ref{eq:ymu}) into Eq.~(\ref{eq:y}), the self-interaction contribution to both $\varepsilon_{\rm D}$ and $p_{\rm D}$ scales as
\begin{equation}
\delta \propto \left(\frac{\sigma}{m_\chi}\right)^{1/2}
\frac{ n_\chi^2}{\overline{m}_\chi^{1/2}}\,,
\label{eq:delta_scaling}
\end{equation}
so that the interaction energy density grows as the square root of the cross section. For fixed $m_\chi$ and $n_\chi$, the two decades in $\sigma/m_\chi$ considered here therefore correspond to only a factor of ten in $\delta$. 
\subsection{Two-fluid hydrostatic equilibrium}
\label{subsec:tov}
The structural properties of the DANSs are determined by solving the two-fluid Tolman-Oppenheimer-Volkoff (TOV) equations for two non-interacting fluids in a static, spherically symmetric, and non-rotating configuration~\cite{Nelson:1803.03266,Ivanytskyi:2109.03801,Leung:2207.02433,Ivanytskyi:2208.03282,Liu:2307.11313}:
\begin{align}
\frac{dp_i}{dr} &= -(p_i + \varepsilon_i)\frac{d\nu}{dr} \:,
\\
\frac{dm_i}{dr}   &= 4\pi r^2 \varepsilon_i \:,
\\
\frac{d\nu}{dr} &= \frac{m + 4\pi r^3 p}{r(r - 2m)} \:,
\label{eq:tovnu}
\end{align}
where $i={\rm N, D}$. The total pressure, energy density, and enclosed gravitational mass within a radius $r$ are $p=p_{\rm N}+p_{\rm D}$, $\varepsilon=\varepsilon_{\rm N}+\varepsilon_{\rm D}$, and $m=m_{\rm N}+m_{\rm D}$. These coupled differential equations are solved with the boundary conditions $m_{\rm N}(0)=m_{\rm D}(0)=0$, $p_{\rm N}(0)=p_c^{{\rm N}}$, and $p_{\rm D}(0)=p_c^{{\rm D}}$ at the center of the star. The integration continues up to the two surfaces, defined by $p_{\rm N}(R_{\rm N})=0$ and $p_{\rm D}(R_{\rm D})=0$. The total gravitational mass and the radius of the system are $M = m_{\rm N}(R_{\rm N}) + m_{\rm D} (R_{\rm D})$ and $R={\rm max}(R_{\rm N}, R_{\rm D})$, respectively. There are therefore DM-core ($R_{\rm D}<R_{\rm N}$) and DM-halo ($R_{\rm D}>R_{\rm N}$) DANS configurations. The second half of this paper is concerned with the quantity
\begin{equation}
\Delta R \;\equiv\; R_{\rm D} - R_{\rm N}\:.
\label{eq:deltaR}
\end{equation}

Because only the baryonic surface is visible to an electromagnetic observer, the radius entering the NICER likelihood is the nucleonic one, and we set $R = R_{\rm N}$ throughout the inference. We emphasize that this concerns only which radius is compared with the data and does not restrict the parameter space: DM-halo configurations are not excluded {\it a priori}, and indeed occur in the posterior at the ten-percent level, as discussed in Sec.~\ref{sec:results}. The tidal deformability, by contrast, responds to the full mass distribution and hence to $R = {\rm max}(R_{\rm N}, R_{\rm D})$. To fix the DM fraction ($f_\chi = M_{\rm D}/M$) inside the DANSs, a root-finding method is used to obtain the required ratio of central pressures. $f_\chi$ is therefore treated as a free parameter in this study, and its prior range is given in Table~\ref{tab:prior_table}.
\subsection{Bayesian inference on the two-fluid DANS system}
\label{subsec:bayes}
A DANS is a complex system, as its two fluids interact with each other only through gravity, and the two-fluid TOV equations must be solved~\cite{Sandin:0809.2942}. The inputs to these equations are the EOSs of both fluids for fixed NM and DM model parameters. To construct the NM and DM EOSs, we use the priors given in Table~\ref{tab:prior_table}.
\begin{table}[htbp]
\centering
\caption{Prior ranges of the NM and DM model parameters used in this study; all priors are uniform.}
\label{tab:prior_table}
\begin{tabular}{lll}
\hline \hline
Parameters &  Prior range & Experimental range \\ \hline
$n_{\rm sat}$ (fm$^{-3}$)       & 0.140 -- 0.170      & 0.140 -- 0.170 \cite{Margueron:1708.06894}     \\ 
$E_{\rm sat}$ (MeV)              & $-16.5$ -- $-15.5$ & -16 -- -15     \cite{Margueron:1708.06894}     \\
$K_{\rm sat}$ (MeV)              & 200 -- 300         & 220 -- 260     \cite{Shlomo:2006.EPJA.30.23, Piekarewicz:0912.5103}  \\
$J_{\rm sat}$ (MeV)              & 25 -- 40           & 28.5 -- 34.9  \cite{Oertel:1610.03361, Li:1304.3368} \\ 
$L_{\rm sat}$ (MeV)              & 40 -- 100          & 30 -- 87       \cite{Oertel:1610.03361, Li:1304.3368} \\ 
$M^*_{\rm sat}$                  & 0.55 -- 0.75       &  ---                                           \\
$\zeta_0$                        & 0.0 -- 4.0         &  ---                                           \\ 
$m_\chi$ (GeV)                   & 0.1 -- 4.0         &  ---                                           \\ 
$f_\chi$                         & 0.0 -- 0.3         &  ---                                           \\
\hline \hline
\end{tabular}
\end{table}

The posterior distribution of the parameter vector $\bm{\theta} = \{n_{\rm sat}, E_{\rm sat}, K_{\rm sat}, J_{\rm sat}, L_{\rm sat}, M^*_{\rm sat}, \zeta_0, m_\chi, f_\chi\}$ follows from Bayes' theorem, $P(\bm{\theta}|d) \propto \mathcal{L}(d|\bm{\theta})\, P(\bm{\theta})$, and is explored by nested sampling with the MultiNest package~\cite{Feroz:0809.3437}. For each sample, the NM EOS is constructed from the six NMPs and $\zeta_0$, matched to the Negele--Vautherin--BPS crust, and rejected if it violates thermodynamic stability, $d\varepsilon>0$, $dp>0$, and $c_s^2>0$, over the whole tabulated range, or causality, $c_s^2<1$, up to the central density $n_{\rm TOV}$ of the maximum-mass star, where $c_s$ is the speed of sound. The DM EOS is constructed from $m_\chi$ alone for a given value of $\sigma/m_\chi$ using Eqs.~(\ref{eq:EOS_DM}) and (\ref{eq:ymu}), and the two-fluid TOV equations are integrated over a grid of $50$ central pressures at the sampled value of $f_\chi$. The total log-likelihood is the sum of five independent terms,
\begin{align}
\ln \mathcal{L} &= \ln \mathcal{L}_{\chi {\rm EFT}}
+ \ln \mathcal{L}_{\rm pQCD}
+ \ln \mathcal{L}_{\rm J0348}
\nonumber \\
&+ \sum_k \ln \mathcal{L}_{\rm NICER,k}
+ \ln \mathcal{L}_{\rm GW170817}.
\label{eq:likelihood}
\end{align}
The first term enforces the $\chi$EFT band for the pressure of pure neutron matter~\cite{Hebeler:1303.4662} at $n_b = 0.08$, $0.12$, and $0.16$~fm$^{-3}$ ($0.5$--$1.0\,n_{\rm sat}$) through a super-Gaussian. For pQCD, we follow the approach of Ref.~\cite{Komoltsev:2111.05350}: the EOS at $n_{\rm TOV}$ must admit a thermodynamically consistent and causal connection to the pQCD EOS at a baryon chemical potential of $2.6$~GeV, and the constraint is marginalized over the renormalization scale $X \in [1,4]$. For the maximum mass, we use the measured mass of PSR~J0348$+$0432, $M= 2.01\pm0.04 \ M_\odot$~\cite{Antoniadis:1304.6875}, instead of a hard cutoff. The fourth term is the sum of the marginalized mass--radius likelihoods of the NICER observations of PSR~J0740$+$6620~\cite{Salmi:2406.14466}, PSR~J0030$+$0451~\cite{Kini:2026.ApJ.1005.201}, PSR J0437$-$4715~\cite{Choudhury:2407.06789}, PSR J0614$-$3329~\cite{Miller:2609.00965}, and PSR J1614$-$2230~\cite{Mauviard:2609.00172}, and the last term is the constraint on the tidal deformability from GW170817~\cite{Abbott:1805.11581}. We repeat the entire inference for three values of the DM self-interaction cross section, $\sigma/m_\chi = 0.1$, $1$, and $10\ {\rm cm^2/g}$, and obtain $\approx 1.4\times10^4$ equally weighted posterior samples in each case.
\begin{figure*}
\centering
\includegraphics[width=1.02\textwidth]{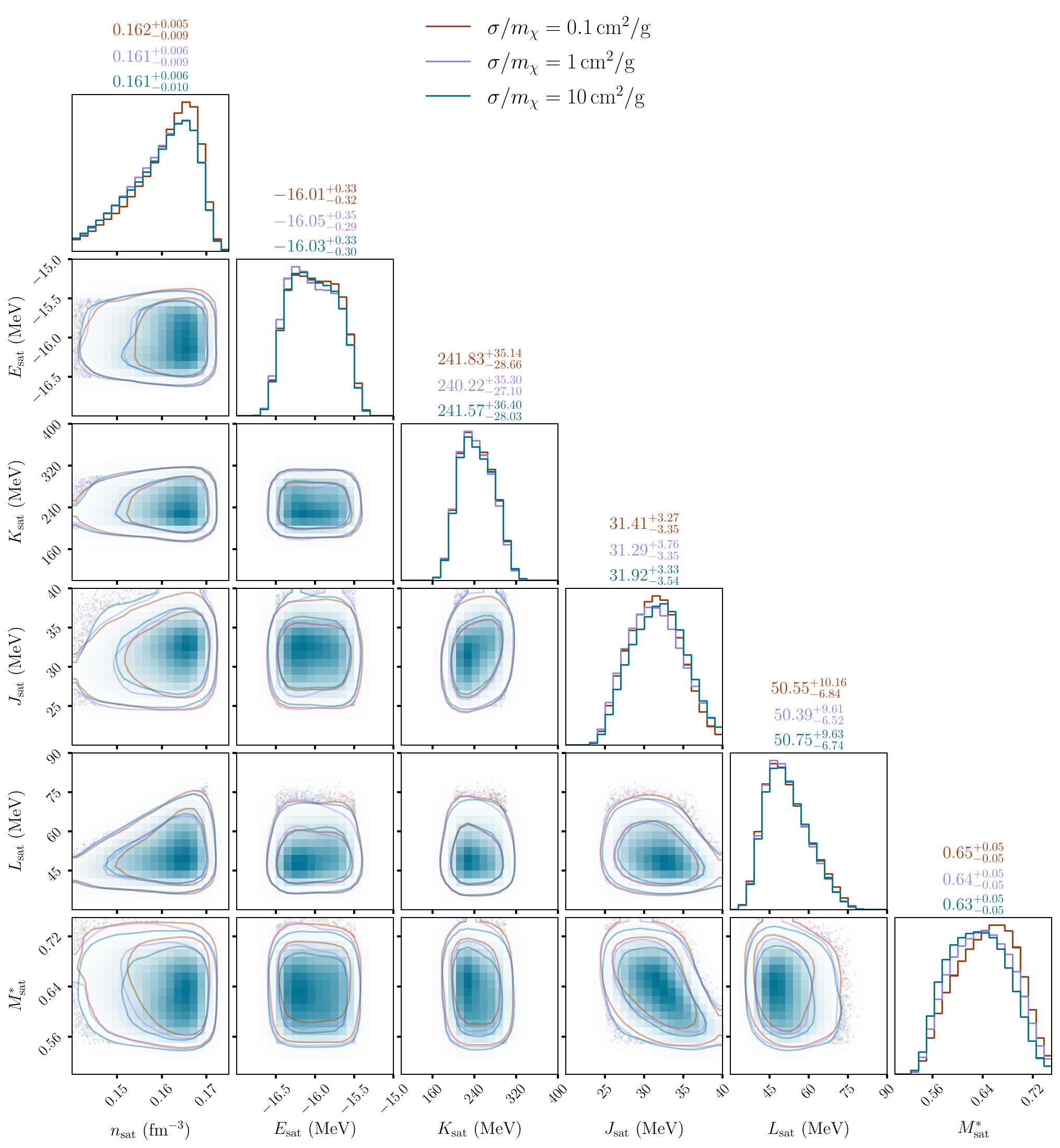}
\caption{Marginalized posterior distributions of the nuclear matter parameters ($n_{\rm sat}$, $E_{\rm sat}$, $K_{\rm sat}$, $J_{\rm sat}$, $L_{\rm sat}$, and $M^*_{\rm sat}$) obtained from the Bayesian analysis of the two-fluid DANS system. The three colors correspond to the dark matter self-interaction strengths $\sigma/m_\chi = 0.1$, $1$, and $10$ cm$^2$/g. The diagonal panels show the one-dimensional marginalized posteriors, with the median and $68\%$ credible interval quoted above each panel, and the off-diagonal panels show the two-dimensional distributions, with contours at the $68\%$ and $95\%$ credible levels.}
\label{fig:corner_nmps}
\end{figure*}
\subsection{The prior-predictive run}
\label{subsec:prior_run}
In addition to the three posteriors, we use a prior-predictive calculation, which serves two distinct purposes. A single set of $5000$ samples is drawn from the priors of Table~\ref{tab:prior_table} and integrated with the same two-fluid solver and the same numerical setup for each of the three cross sections. First, in Sec.~\ref{subsec:RD_data_favour}, it supplies the denominator of the Bayes factor for halo configurations.

Second, because the parameter draws are shared between the three runs, sample $i$ at $\sigma/m_\chi = 10\ {\rm cm^2/g}$ and sample $i$ at $0.1\ {\rm cm^2/g}$ differ only in the cross section; the nuclear parameters, $m_\chi$, and $f_\chi$ are identical by construction. Differencing the two runs sample by sample therefore cancels these parameters exactly, rather than averaging over them, and yields a paired estimate of the effect of $\sigma/m_\chi$ alone that no comparison of independent posterior draws can match. This is used in Sec.~\ref{subsec:RD_self_interaction}.
\section{Results}
\label{sec:results}
\subsection{Nuclear matter parameters}
\label{subsec:RD_NMPs}
Figure~\ref{fig:corner_nmps} shows the posterior distributions of the nuclear matter parameters for the three values of $\sigma/m_\chi$. The recovered values are nearly identical in all three cases. For $\sigma/m_\chi = 1$ cm$^2$/g, we find $n_{\rm sat} = 0.161^{+0.006}_{-0.009}$ fm$^{-3}$, $E_{\rm sat}= -16.05_{-0.29}^{+0.35}$ MeV, $K_{\rm sat} = 240.22_{-27.10}^{+35.30}$ MeV, $J_{\rm sat} = 31.29_{-3.35}^{+3.76}$ MeV, $L_{\rm sat} = 50.39_{-6.52}^{+9.61}$ MeV, and $M^*_{\rm sat} = 0.64\pm0.05$, with $68\%$ credible intervals (CIs) that overlap almost completely with those for $\sigma/m_\chi=0.1$ and $10$ cm$^2$/g; the medians differ by only a small fraction of the quoted uncertainties. The posterior of $n_{\rm sat}$ peaks in the upper part of its prior, and that of $L_{\rm sat}$ is pushed towards the lower edge. $E_{\rm sat}$, $J_{\rm sat}$, and $M^*_{\rm sat}$ are well constrained within their prior ranges, and $K_{\rm sat}$ favors values in the lower part of its prior. The two-dimensional contours show the correlations between each pair of parameters.

The three sets of contours lie on top of one another, which shows that the inferred NMPs are essentially insensitive to the dark matter self-interaction strength over the range considered ($\sigma/m_\chi = 0.1$ to $10$ cm$^2$/g). The origin of this insensitivity lies in the structure of the model rather than in the statistics, and it can be traced through four successive suppressions.
\begin{itemize}
\item The cross section is not an independent degree of freedom of the dark sector; it enters only through the Born-approximation relation, Eq.~(\ref{eq:ymu}), which fixes the interaction parameter $y$ once $m_\chi$ has been drawn. Varying the cross section therefore does not open a new direction in parameter space; it only reparametrizes the map between $m_\chi$ and $y$.

\item The cross section reaches $\varepsilon_{\rm D}$ and $p_{\rm D}$ only through the interaction term
$\delta$, which by Eq.~(\ref{eq:delta_scaling}) grows as $(\sigma/m_\chi)^{1/2}$: two decades
in the cross section correspond to a factor of ten in $\delta$ and of only $10^{1/4} \simeq 3.2$ in $y$ itself, whereas $m_\chi$ alone changes $y$ by a factor of $16$ across the prior of Table~\ref{tab:prior_table} within a single inference. The three runs therefore sample largely overlapping ranges of $y$, as is evident in Fig.~\ref{fig:corner_dm}.

\item Whatever residual difference survives is weighted by the dark matter mass fraction, which the data confine to $f_\chi < 0.13$, $0.16$, and $0.19$ ($90\%$ upper limits) for $\sigma/m_\chi = 0.1$, $1$, and $10$~cm$^2$/g. Isolating the dependence at fixed $f_\chi$, as we do in Sec.~\ref{subsec:RD_sensitivity}, shows that the full range of $\sigma/m_\chi$ shifts the posterior $R_{1.4}$ by an amount within $\sim0.015$~km, and by at most $\sim0.1$~km near $M_{\rm max}$, even in the region most favorable to it.

\item The nuclear sector is anchored in all three cases by the same likelihood (Eq.~(\ref{eq:likelihood})), with the same $\chi$EFT band, pQCD limit, and NICER and GW170817 measurements. Within the posterior, the residual shift in radius is one to two orders of magnitude below the $\sim1$~km precision of the NICER radii, so the likelihood barely distinguishes the three cross sections and the NMPs recover the same posterior in each case. As we show in Sec.~\ref{subsec:RD_self_interaction}, this is not because the self-interaction is intrinsically weak, but because the data disfavor the light-$m_\chi$ region where its effect is large.
\end{itemize}
The parameter $\zeta_0$, which controls the stiffness of the nuclear EOS at supra-saturation density, is constrained to the lower part of its prior, although it is sampled uniformly over $[0,4]$: its posterior median is $0.58^{+0.83}_{-0.37}$. The mechanism is transparent: $\zeta_0$ softens the EOS at supra-saturation density, and we find a correlation coefficient of $-0.22$ between $\zeta_0$ and $M_{\rm max}$, with the median maximum mass falling from $2.08\, M_\odot$ for $\zeta_0<0.5$ to $2.06\, M_\odot$ for $1.0<\zeta_0<1.5$. The drop is small because the other NMPs and the dark matter fraction adjust to keep $M_{\rm max}$ close to the masses of the heavy pulsars, but large values of $\zeta_0$, which would soften the EOS too much, are disfavored.
\begin{figure}
\centering
\includegraphics[width=\columnwidth]{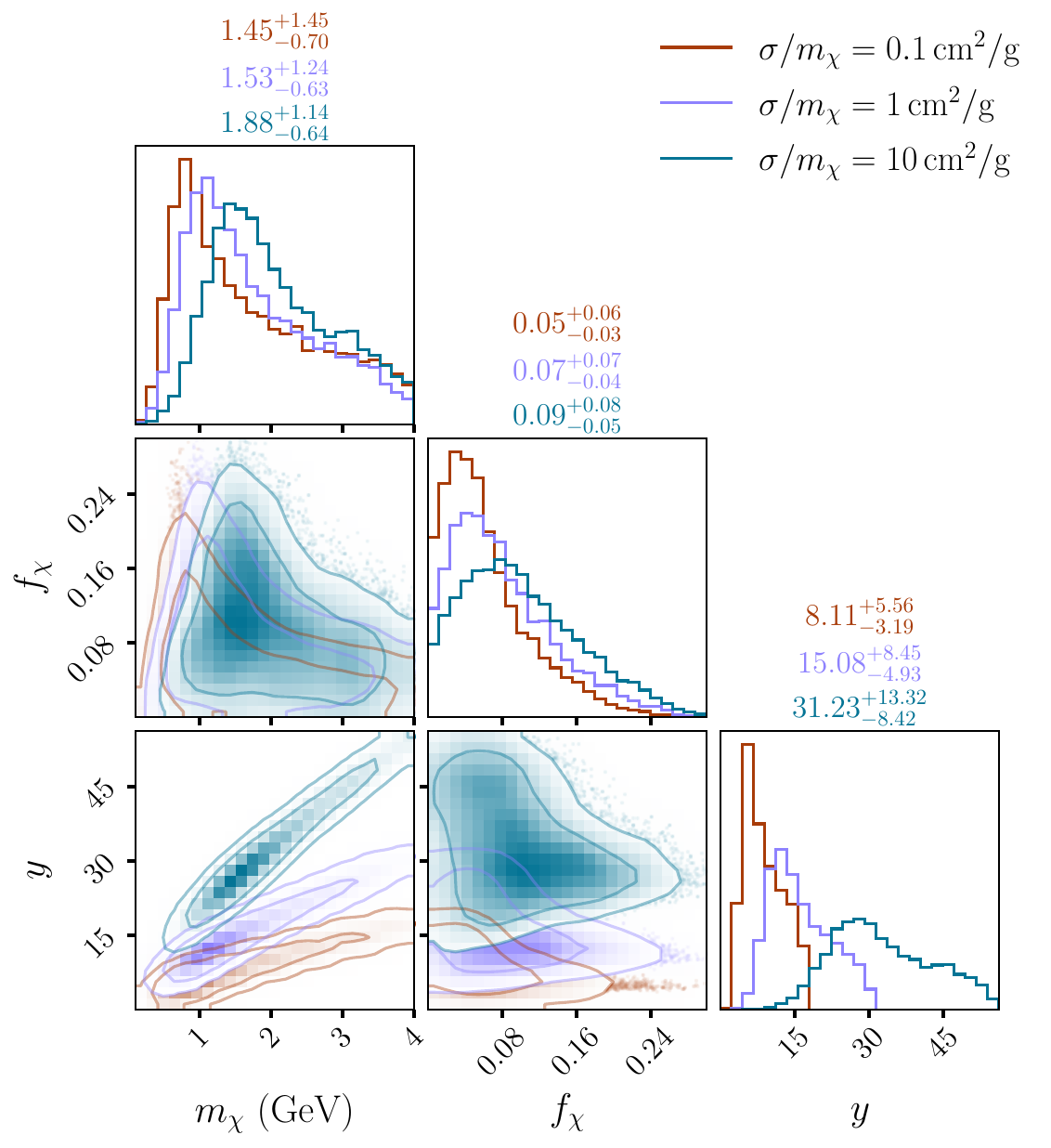}
\caption{Marginalized posterior distributions of $m_\chi$, $f_\chi$, and the interaction parameter $y$, obtained from Eq.~(\ref{eq:ymu}), for the DANS system at the three dark matter cross sections. The quoted values above the diagonal panels are medians with $68\%$ credible intervals.}
\label{fig:corner_dm}
\end{figure}
\begin{figure}[h]
\centering
\includegraphics[width=0.5\textwidth]{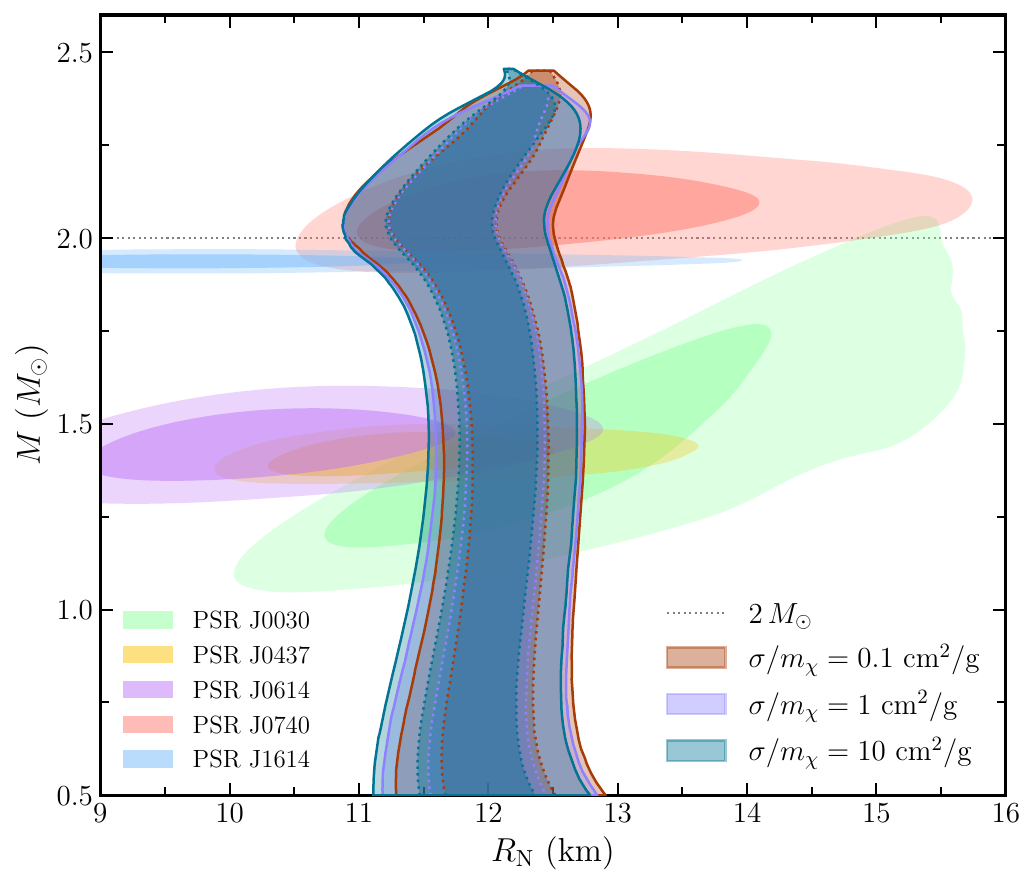}\\
\caption{The $M$--$R_{\rm N}$ relations obtained for the two-fluid DANS system, with $68\%$ and $95\%$ credible regions. The bands are drawn up to the largest mass reached by any posterior configuration. The observational constraints used in the inference are also shown: the $2\, M_\odot$ limit and the NICER measurements of PSR~J0740$+$6620~\cite{Salmi:2406.14466}, PSR~J0030$+$0451~\cite{Kini:2026.ApJ.1005.201}, PSR J0437$-$4715~\cite{Choudhury:2407.06789}, PSR J0614$-$3329~\cite{Miller:2609.00965}, and PSR J1614$-$2230~\cite{Mauviard:2609.00172}.}
\label{fig:mr_posterior}
\end{figure}
\subsection{Dark matter parameters}
\label{subsec:RD_DMPs}
In Fig.~\ref{fig:corner_dm}, we show the posteriors of the dark sector parameters: the particle mass $m_\chi$, the mass fraction $f_\chi$, and the interaction strength $y$ that follows from Eq.~(\ref{eq:ymu}). The dark matter mass increases slightly as the cross section grows, from $m_\chi = 1.45^{+1.45}_{-0.70}$ through $1.53^{+1.24}_{-0.63}$ to $1.88^{+1.14}_{-0.64}$~GeV, and the corresponding interaction parameter from $y = 8.1$ to $15.1$ and $31.2$. Once $\sigma/m_\chi$ is fixed, $y$ is simply a function of $m_\chi$, so the narrow diagonal bands in the $m_\chi$--$y$ panels show that function rather than a constraint found by the data. The three runs are offset from one another by the fixed factor $(\sigma/m_\chi)^{1/4}$, which is only $(10/0.1)^{1/4}\sim3.2$ across the whole range of $\sigma/m_\chi$ considered, whereas $m_\chi$ alone changes $y$ by a factor of $(4.0/0.1)^{3/4}\sim16$ within a single run.

The shift in $m_\chi$ is not a uniform displacement but a loss of the light-mass region. The posterior probability of $m_\chi < 1$~GeV falls from $31.6\%$ to $21.3\%$ and $7.4\%$ across the three runs, while the high-mass tail is nearly unchanged. The reason lies in Eq.~(\ref{eq:delta_scaling}). At fixed $\sigma/m_\chi$, the interaction term scales as $\overline{m}_\chi^{-1/2}$, so the self-interaction acts most strongly on light particles, and a stronger cross section turns the light dark component into an extended, stiff cloud, a halo configuration, which is penalized by the data. We discuss halo and core configurations further in Secs.~\ref{subsec:RD_core_halo} and \ref{subsec:RD_self_interaction}.

The mass fraction follows a similar trend. Comparing the three runs, the median increases with the cross section, from $0.05$ to $0.07$ and $0.09$. A non-zero fraction is required by the pulsars with the smallest inferred radii, PSR~J0437$-$4715, PSR~J0614$-$3329, and PSR~J1614$-$2230, and by GW170817, all of which favor more compact stars: a dark core contracts the baryonic star and lowers $\Lambda_{1.4}$, while PSR~J0740$+$6620 and PSR~J0030$+$0451 pull in the opposite direction. The $90\%$ upper bound, $f_\chi < 0.13$, $0.16$, and $0.19$, loosens with the cross section. It is set by the heavy pulsars, PSR~J0740$+$6620, PSR~J1614$-$2230, and PSR~J0348$+$0432, because dark matter lowers $M_{\rm max}$. A stronger self-interaction stiffens the dark fluid, so that a given dark fraction costs less maximum mass, and correspondingly more dark matter is allowed.
\begin{table}
\centering
\caption{Median values with $68\%$ credible intervals of the DANS observables for each dark matter self-interaction strength. Radii are those of the baryonic surface, $R_{\rm N}$.}
\label{tab:medians}
{\renewcommand{\arraystretch}{1.6}
\begin{tabular}{lccc}
\hline\hline
Quantity & \shortstack{$\sigma/m_\chi = 0.1$ \\ cm$^2$/g} & \shortstack{$\sigma/m_\chi = 1$ \\ cm$^2$/g} & \shortstack{$\sigma/m_\chi = 10$ \\ cm$^2$/g} \\
\hline
$M_{\rm max}$ ($M_\odot$) & $2.07^{+0.07}_{-0.05}$ & $2.07^{+0.07}_{-0.05}$ & $2.07^{+0.07}_{-0.05}$ \\
$R_{\rm max}$ (km)        & $11.08^{+0.34}_{-0.33}$ & $11.07^{+0.34}_{-0.32}$ & $11.04^{+0.34}_{-0.31}$ \\
$R_{1.4}$ (km)            & $12.16^{+0.30}_{-0.28}$ & $12.13^{+0.31}_{-0.29}$ & $12.06^{+0.31}_{-0.29}$ \\
$\Lambda_{1.4}$           & $390^{+80}_{-63}$ & $385^{+84}_{-64}$ & $381^{+87}_{-63}$ \\
\hline\hline
\end{tabular}}
\end{table}
\begin{figure*}
\centering
\includegraphics[width=0.9\textwidth]{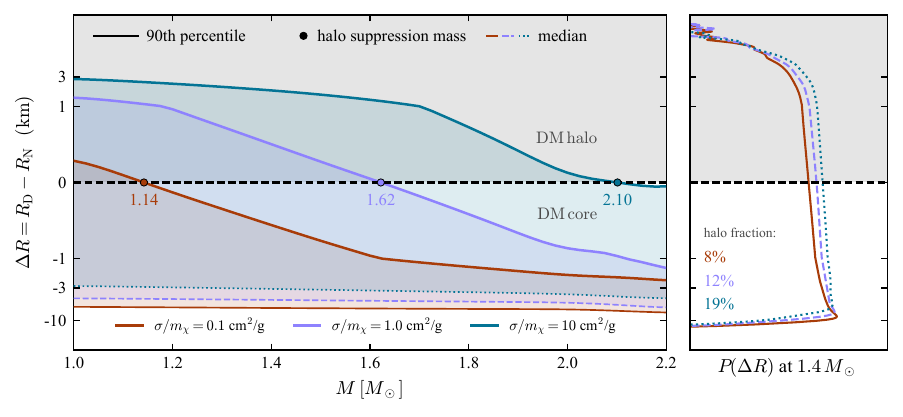}
\caption{The difference between the dark and baryonic surfaces, $\Delta R = R_{\rm D} - R_{\rm N}$, as a function of stellar mass along the stable branch, for the three dark matter self-interaction strengths
$\sigma/m_\chi = 0.1$, $1$, and $10$ cm$^2$/g. Solid curves are the $90$th percentile of the posterior at each mass, dotted/dashed/solid curves the corresponding medians, and the shaded bands span the region between the median and the $90$th percentile. The horizontal dashed line at $\Delta R = 0$ separates DM-core ($\Delta R < 0$) from DM-halo ($\Delta R > 0$) configurations; the grey region above it is the halo side. Filled circles mark the halo suppression mass, the stellar mass at which the $90$th percentile crosses zero, i.e.\ the heaviest star for which the upper percentile of the posterior is still a halo. The right panel shows the distribution of $\Delta R$ at $1.4\, M_\odot$ on a logarithmic scale, with the halo fraction quoted in the panel.}
\label{fig:deltaR}
\end{figure*}
\subsection{Global observables}
\label{subsec:RD_global_obs}
Figure~\ref{fig:mr_posterior} shows the mass--radius relations together with the observational constraints used in the inference. The credible bands of the three cases nearly coincide over the entire mass range and lie comfortably within the NICER contours, which confirms that the inference reproduces the data on which it was conditioned. The medians with $68\%$ credible intervals of $M_{\rm max}$, $R_{\rm max}$, $R_{1.4}$, and $\Lambda_{1.4}$ are listed in Table~\ref{tab:medians}.

For every observable, the three cross sections give almost the same result. The medians of $M_{\rm max}$ agree to $0.001\,M_\odot$, while $R_{1.4}$ decreases by $0.10$~km and $\Lambda_{1.4}$ by $9$ units from $\sigma/m_\chi = 0.1$ to $10\ {\rm cm^2/g}$, reflecting the slightly larger dark fraction at the higher cross section. The global observables are therefore stable across two decades of uncertainty in the self-interaction. Since no hard cut on the maximum mass is imposed, the lower tail of the $M_{\rm max}$ posterior is informative: a small fraction of the posterior has $M_{\rm max}<2\,M_\odot$, which is permitted by the measured mass of PSR~J0348$+$0432, $2.01\pm0.04 \ M_\odot$. In the upper tail, the $97.5$th percentile lies at $2.22\,M_\odot$, and very few samples reach $2.3\,M_\odot$. The nuclear EOS ensemble that satisfies $\chi$EFT, pQCD, and the astrophysical data therefore comfortably supports massive pulsars, in the same way for every cross section. The same nuclear matter parameters and global observables are obtained when $\sigma/m_\chi$ is treated as a free parameter and marginalized over (Appendix~\ref{app:sigma_free}).
\subsection{Core and halo configurations}
\label{subsec:RD_core_halo}
The previous sections show that the global observables are insensitive to the self-interaction; the internal geometry of the admixed star, however, is not. To explore this, we classify every posterior configuration as a DM core or a DM halo according to the sign of $\Delta R$, Eq.~(\ref{eq:deltaR}), obtained by interpolating $R_{\rm N}$ and $R_{\rm D}$ along the stable branch of each sample, and show it as a function of stellar mass in Fig.~\ref{fig:deltaR}. In all three cases, the posterior median of $\Delta R$ is negative at every mass between $0.5$ and $2.2\,M_\odot$, i.e.\ the typical posterior sample is a DM core. For a $1.4\,M_\odot$ star, the median dark surface lies $\Delta R \simeq -6.3$, $-4.7$, and $-3.1$~km inside the baryonic one, corresponding to a dark core of radius $R_{\rm D}\simeq5.9$, $7.4$, and $8.9$~km within a star of $R_{\rm N}\simeq12.1$~km. Unlike the global observables of Secs.~\ref{subsec:RD_NMPs}--\ref{subsec:RD_global_obs}, however, the size of the dark core changes monotonically with the cross section, that means a stronger self-interaction stiffens the dark fluid, which then extends further out at a given mass fraction. 

In the upper tail, however, the $90$th percentile curves separate cleanly, and each crosses $\Delta R = 0$ at a well-defined mass, which we call the ``halo suppression mass": the heaviest star for which the upper percentile of the posterior still extends beyond the baryonic surface. It rises monotonically with the cross section,
\begin{equation}
M_{\rm suppression} = 1.14,\ 1.62,\ 2.10\ M_\odot
\label{eq:suppression}
\end{equation}
for $\sigma/m_\chi = 0.1$, $1$, and $10\ {\rm cm^2/g}$. A shift of $\sim1.0\, M_\odot$ across two decades in the cross section is a large effect. It arises because a stronger self-interaction stiffens the dark fluid, which then spreads further out at a given mass fraction, so that increasingly heavy stars can still support a halo in the tail of the posterior. The right panel shows the distribution of $\Delta R$ for a $1.4\, M_\odot$ star; the halo fractions are $8$, $12$, and $19\%$ for the three cross sections. For $2.0\,M_\odot$, the corresponding values are $3$, $6$, and $10\%$. The suppression at higher mass reflects the deeper gravitational well, which pulls the dark component inside the baryonic star.
\begin{figure}
\centering
\includegraphics[width=1.0\columnwidth]{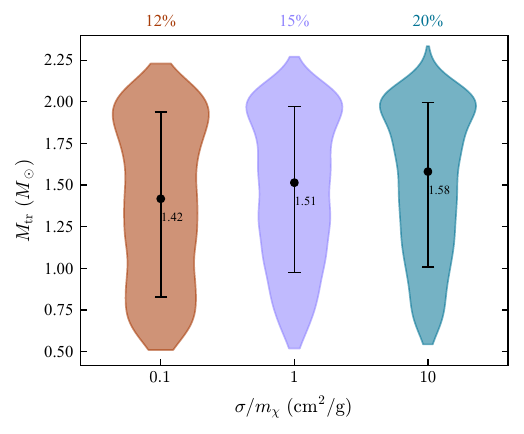}
\caption{Distribution of the core-to-halo transition mass $M_{\rm tr}$, the heaviest star along a posterior sequence that still satisfies $R_{\rm D} > R_{\rm N}$, for the three self-interaction strengths. Violins are kernel density estimates over the sequences that undergo a transition and span the full range of $M_{\rm tr}$. The dot is the median (value quoted) and the capped bar the $16$th--$84$th percentile range. The percentage above each violin is the fraction of posterior sequences that cross $\Delta R = 0$ along the stable branch above $0.5\,M_\odot$.}
\label{fig:transition}
\end{figure}

To understand the change in geometry further, we treat each posterior sample as a sequence and follow how $\Delta R$ falls with $M$. A sample that is a halo at low mass and a core at high mass has a well-defined transition mass $M_{\rm tr}$, shown in Fig.~\ref{fig:transition}. Similar configuration classes, named ``Core-Halo'', were introduced in the agnostic two-fluid study of Ref.~\cite{karan:2607.03840}. We find $M_{\rm tr} = 1.42^{+0.52}_{-0.59}$, $1.51^{+0.46}_{-0.54}$, and $1.58^{+0.42}_{-0.57}\,M_\odot$, occurring in $12\%$, $15\%$, and $20\%$ of the posterior sequences. Both the location and the incidence increase monotonically with the cross section, but the shift of the median is small compared with the width of the distribution; $M_{\rm tr}$ is therefore a useful descriptor of the geometry rather than a discriminator between cross sections.
\subsection{Where the sensitivity lies}
\label{subsec:RD_sensitivity}
The results of Secs.~\ref{subsec:RD_NMPs}--\ref{subsec:RD_global_obs} raise an obvious question: if the self-interaction term $\delta$ changes by a factor of ten across the range of cross sections considered, why is nothing visible in the observables? The marginalized posteriors cannot answer this, because they average over $f_\chi$ and $m_\chi$, both of which shift between the runs. We therefore compare the extreme inferences at fixed DM mass fraction and at every stellar mass in Fig.~\ref{fig:residual}.
\begin{figure}
\centering
\includegraphics[width=1.0\columnwidth]{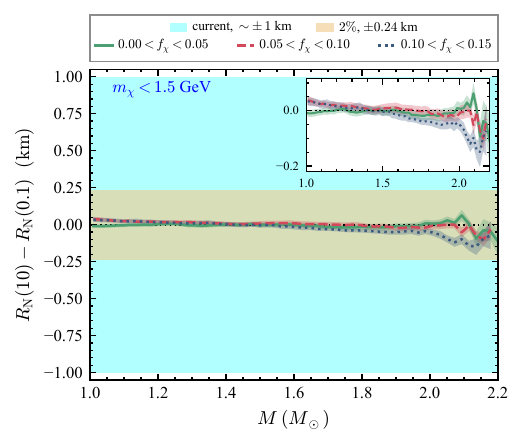}
\caption{The displacement of the baryonic radius between the extreme cross sections, $R_{\rm N}(10\ {\rm cm^2/g}) - R_{\rm N}(0.1\ {\rm cm^2/g})$, at fixed DM mass fraction and restricted to $m_\chi < 1.5$~GeV, where the dark sector is most influential. Curves are medians in three $f_\chi$ bins, and the bands are the errors on the median. The cyan and orange strips mark the present radius precision of order $1$~km and the $\pm 0.24$~km that a $2\%$ measurement of a $\sim12$~km star would deliver, respectively.}
\label{fig:residual}
\end{figure}

Figure~\ref{fig:residual} shows the displacement of the baryonic radius between $\sigma/m_\chi = 0.1$ and $10\ {\rm cm^2/g}$ in three bins of $f_\chi$, restricted to $m_\chi<1.5$~GeV, where the dark sector perturbs the star most strongly (see Fig.~\ref{fig:prior}a). This cut keeps the whole region in which the paired response of Sec.~\ref{subsec:RD_self_interaction} is large, while retaining enough samples in every bin. At $1.4\,M_\odot$, the displacement is $+0.002 \pm 0.015$, $+0.006 \pm 0.014$, and $+0.003 \pm 0.015$~km for $0<f_\chi<0.05$, $0.05<f_\chi<0.10$, and $0.10<f_\chi<0.15$. It remains below $0.05$km in every $f_\chi$ bin at all masses up to about $1.9\,M_\odot$ and grows only close to the maximum mass, reaching $-0.11$~km at $2.1\,M_\odot$, while in the middle bin it reaches $-0.05$~km, which is not yet significant. Since the displacement is largest there, massive stars with a size able dark fraction are the most informative targets.

This displacement, however, is not measurable at present. Radius measurements by NICER carry an uncertainty of order $1$~km, so even the largest shift in Fig.~\ref{fig:residual} amounts to about a tenth of a standard deviation, and it lies well inside the $\pm0.24$~km that a future $2\%$ measurement would deliver. Distinguishing $\sigma/m_\chi = 0.1$ from $10\ {\rm cm^2/g}$ through the posterior radius would therefore require a precision well below $0.1$~km, far beyond current capability. Neutron star observables therefore cannot at present discriminate between SIDM cross sections in the range permitted by cluster and dwarf-galaxy observations. One caveat applies to Fig.~\ref{fig:residual}: the posteriors are independent sets of draws whose $m_\chi$ distributions differ between runs (Sec.~\ref{subsec:RD_DMPs}), so what is measured here is how far an observable moves across the posterior rather than its response with all other parameters held fixed. We provide an estimate free of this limitation in Sec.~\ref{subsec:RD_self_interaction}.
\begin{figure*}
    \centering
    \includegraphics[width=1.0\textwidth]{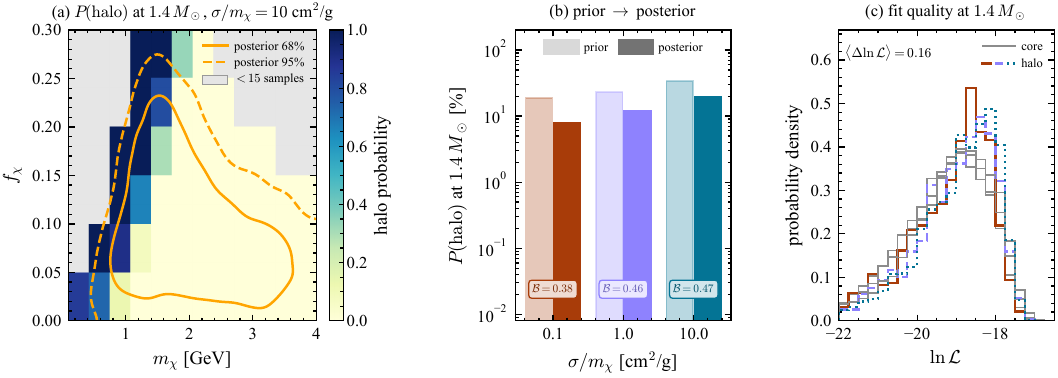}
    \caption{Evidence on DM-halo configurations. (a) Probability that a configuration of $1.4\,M_\odot$ is a halo, as a function of $m_\chi$ and $f_\chi$, for $\sigma/m_\chi = 10$ cm$^2$/g; grey cells contain fewer than $15$ posterior samples and are excluded, and the orange curves are the $68\%$ and $95\%$ credible contours of the
    posterior itself. (b) Halo probability at $1.4\,M_\odot$ before (light) and after (dark) the astrophysical data, on a logarithmic scale; the prior values are obtained by solving the two-fluid TOV equations for draws from the prior, and the Bayes factor $\mathcal{B}_{\rm halo}$ is quoted for each case. (c) Distribution of the log-likelihood for core (grey) and halo (colored by $\sigma/m_\chi$) posterior samples at $1.4\,M_\odot$, each distribution normalized independently.}
    \label{fig:halo_evidence}
\end{figure*}
\subsection{Do the data favor halo configurations?}
\label{subsec:RD_data_favour}
A halo fraction of order ten percent is not informative by itself, because it depends on how common halo-producing parameters are in the prior. The quantity that answers the question is the change from prior to posterior odds,
\begin{equation}
\mathcal{B}_{\rm halo} =
\frac{P_{\rm post}({\rm halo})/P_{\rm post}({\rm core})}
{P_{\rm prior}({\rm halo})/P_{\rm prior}({\rm core})} \:,
\label{eq:bayes_halo}
\end{equation}
with $\mathcal{B}_{\rm halo}>1$ indicating that the data prefer halos and $\mathcal{B}_{\rm halo}<1$ that they disfavor them. The denominator is evaluated on the prior-predictive run of Sec.~\ref{subsec:prior_run}, with the halo probability read off the resulting sequences in exactly the same way as for the posterior. Both probabilities are conditional on the same event, namely that a configuration of the given mass exists on the stable branch, so the odds entering Eq.~(\ref{eq:bayes_halo}) are comparable. The results are given in Table~\ref{tab:halo} and shown in Fig.~\ref{fig:halo_evidence}.
\begin{table}
\centering
\caption{Prior and posterior probabilities of a DM-halo configuration at $1.4$ and $2.0\,M_\odot$, and the resulting Bayes factor $\mathcal{B}_{\rm halo}$. The prior values are measured directly by solving the two-fluid TOV equations for draws from the prior; $N_{\rm prior}$ is the number of prior sequences that admit a configuration of that mass. $P_{\rm post}$ is the fraction of posterior samples whose stable branch reaches the given mass and for which $R_{\rm D} > R_{\rm N}$ there; $P_{\rm prior}$ is defined in the same way on the prior-predictive draws, so that both are conditional on the existence of a configuration of that mass.}
\label{tab:halo}
{\renewcommand{\arraystretch}{1.3}
\begin{tabular}{lccccc}
\hline\hline
$\sigma/m_\chi$ (cm$^2$/g) & $M$ ($M_\odot$) & $N_{\rm prior}$ & $P_{\rm prior}$ (\%) & $P_{\rm post}$ (\%) & $\mathcal{B}_{\rm halo}$ \\
\hline
$0.1$  & $1.4$ & $3479$ & $19$  & $8$ &  $0.38$ \\
$1$    & $1.4$ & $3938$ & $23$  & $12$ & $0.46$ \\
$10$   & $1.4$ & $4250$ & $34$  & $19$ & $0.47$ \\
\hline
$0.1$  & $2.0$ & $1029$ & $34$  & $3$ & $0.06$ \\
$1$    & $2.0$ & $1233$ & $42$  & $6$ & $0.09$ \\
$10$   & $2.0$ & $1571$ & $52$  & $10$ & $0.11$ \\
\hline\hline
\end{tabular}}
\end{table}

In every case, the data reduce the odds of a halo configuration. At $1.4 \ M_\odot$ the reduction is moderate, with $\mathcal{B}_{\rm halo} = 0.38$, $0.46$, and $0.47$, while for $2 \ M_\odot$ stars the suppression is much stronger, by a factor of ten or more. The mechanism is visible in panel (a): halo configurations require both light $m_\chi$ and an appreciable $f_\chi$, a combination that lowers the maximum mass and is therefore penalized by the heavy pulsars. As a result, the posterior credible regions extend only partially into that corner of parameter space.

Panel (c) compares the fit quality of the halo and core samples within each posterior through the difference of their mean log-likelihoods,
\begin{align}
    \Delta\ln\mathcal{L} = \langle\ln\mathcal{L}\rangle_{\rm halo} - \langle\ln\mathcal{L}\rangle_{\rm core},
\end{align}
evaluated at fixed $\sigma/m_\chi$. For a $1.4 \ M_\odot$ star, $\Delta\ln\mathcal{L} = 0.15$, $0.21$, and $0.12$ for $\sigma/m_\chi = 0.1$, $1$, and $10\ {\rm cm^2/g}$, respectively, with an average of $0.16$. The halo configurations that survive in the posterior therefore fit the data as well as the core configurations, or even slightly better. Most of the parameter volume that produces halos for the light $m_\chi$ with an appreciable $f_\chi$, lowers the maximum mass below what the heavy pulsars allow and is removed by the data. It is worth noting that both $P_{\rm prior}$ and $P_{\rm post}$ increase with $\sigma/m_\chi$: the cross section controls how many halo configurations the model can produce, and the data preserve that ordering. The penalty they apply is nearly independent of the cross section at $1.4\,M_\odot$, where $\mathcal{B}_{\rm halo}$ ranges only from $0.38$ to $0.47$, but weakens noticeably at $2\,M_\odot$, from $0.06$ to $0.11$, because a stiffer dark fluid costs less maximum mass and is therefore less constrained by the heavy pulsars.
\begin{figure*}
\centering
\includegraphics[width=1.0\textwidth]{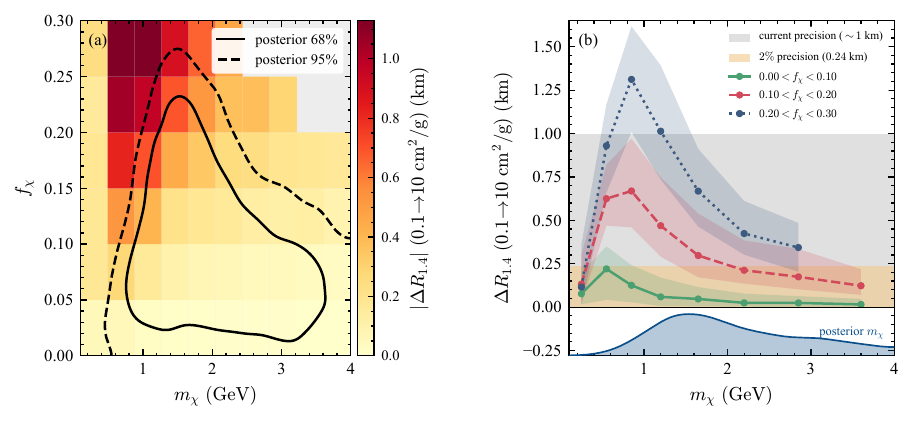}
\caption{(a) Median paired displacement $|\Delta R_{1.4}|$ between $\sigma/m_\chi = 0.1$ and $10$ cm$^2$/g over the $(m_\chi, f_\chi)$ plane, obtained from the prior-predictive run of Sec.~\ref{subsec:prior_run}, in which the same parameter draws were integrated at all three cross sections; grey cells hold fewer than $12$ paired draws. Black curves are the $68\%$ and $95\%$ credible contours of the posterior at $10\ {\rm cm^2/g}$. (b) The same displacement against $m_\chi$ in three $f_\chi$ bins, with $16$th--$84$th percentile bands; the horizontal strips mark the present radius precision and a $2\%$ measurement, as in Fig.~\ref{fig:residual}. The blue profile below the axis is the posterior distribution of $m_\chi$ at $10$ cm$^2$/g, on an arbitrary vertical scale.}
\label{fig:prior}
\end{figure*}
\subsection{Where the self-interaction could be seen}
\label{subsec:RD_self_interaction}
In Sec.~\ref{subsec:RD_sensitivity}, we compared the three separate inferences and noted the caveat that their $m_\chi$ samples differ between cross sections. The prior-predictive run does not have this problem: there, the same parameter draws were integrated three times, once for each cross section, so that two runs can be compared draw by draw. We compute the quantity $\Delta R_{1.4} = R_{\rm N}(10) - R_{\rm N}(0.1)$ for a $1.4\,M_\odot$ star at the two extreme cross sections, shown in Fig.~\ref{fig:prior}.

Paired in this way, the effect of the cross section is far from negligible. Going from $0.1$ to $10\ {\rm cm^2/g}$ makes the star larger by about $0.16$~km; in $40\%$ of the draws the shift exceeds the $0.24$~km that a $2\%$ radius measurement could resolve. The effect is concentrated at light dark matter region, as the right panel of Fig.~\ref{fig:prior} shows. It is largest for $m_\chi$ just below $1$~GeV, where it reaches about $0.7$~km and exceeds a kilometer for the largest dark fractions, and it fades for heavier particles. 

For very light particles, it becomes small again, because the dark matter is then too dilute to affect the baryonic star. The paired estimate keeps everything else fixed and shows what the cross section does on its own: a stiffer dark fluid pushes outwards and supports more mass, the same tendency that raises the halo fraction in Sec.~\ref{subsec:RD_core_halo}. The posterior comparison, in contrast, involves three separate inferences, and as the cross section grows, the posterior moves away from the light-$m_\chi$ region where the effect resides.

The important point of Fig.~\ref{fig:prior} is therefore not the size of the effect but where it sits. The bright region in the left panel and the peak in the right panel both lie at $m_\chi \lesssim 1.5$~GeV, a region the posterior largely abandons as the cross section increases (Sec.~\ref{subsec:RD_DMPs}). The self-interaction is not weak, however it changes the star noticeably for light dark matter. The data disfavor exactly these light-dark-matter configurations, the same ones that form halos, so its effect does not show up in the posterior. The uncertainty in $\sigma/m_\chi$ is therefore harmless for bounds on the nuclear EOS, but not for bounds on the dark sector. The upper bound on $f_\chi$ and the support for sub-GeV dark matter both depend on the cross section.
\begin{figure*}
\centering
\includegraphics[width=1.0\textwidth]{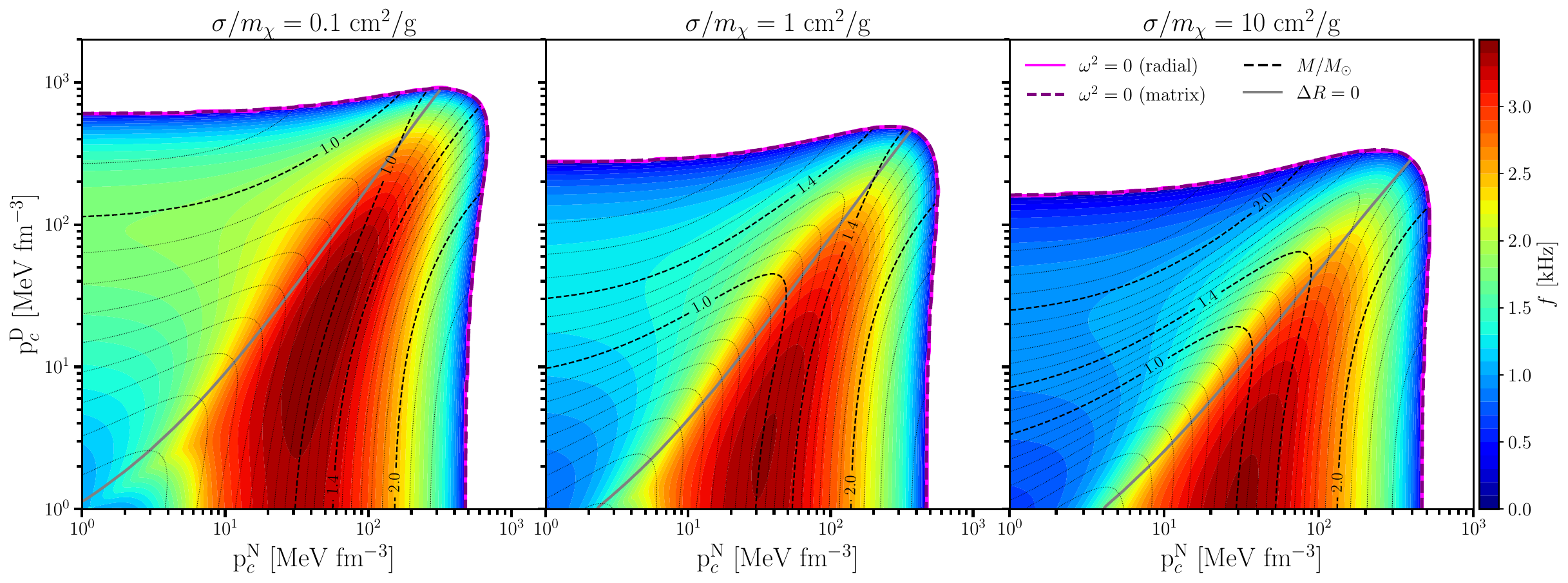}
\caption{Fundamental radial oscillation frequency $f$ of the two-fluid configurations in the plane of the two central pressures $p_c^{\rm N}$ and $p_c^{\rm D}$, for $\sigma/m_\chi = 0.1$, $1$, and $10$ cm$^2$/g (left to right), evaluated for the median posterior nuclear parameters and $m_\chi$ of each case; $f_\chi$ varies across the plane. The magenta solid curve marks the stability boundary $\omega^2 = 0$ obtained from the radial oscillation equations, and the purple dashed curve the same boundary obtained from the two-parameter turning-point (matrix) criterion; the two agree closely. Black dashed curves are contours of constant gravitational mass $M/M_\odot$, and the grey solid curve is the locus $\Delta R = R_{\rm D} - R_{\rm N} = 0$, which separates DM-core configurations (below) from DM-halo configurations (above).}
\label{fig:stability}
\end{figure*}
\subsection{Radial stability}
\label{subsec:stability}
Figure~\ref{fig:stability} shows the fundamental radial oscillation frequency in the plane of the two central pressures for the median posterior EOS of each case, built from the posterior medians of the nuclear parameters and of $m_\chi$; the dark matter fraction is not fixed but varies across the plane. For a two-fluid star, the classical turning-point criterion must be generalized, since the equilibrium sequence is a two-parameter family. Therefore we determine the stability boundary both from the vanishing of the squared frequency of the fundamental radial mode and from the two-parameter (matrix) generalization of the turning-point theorem \cite{Kain:2102.08257, Hippert:2211.08590}. The two criteria agree closely across the whole plane, which provides a useful internal check of the calculation.

The stable region is bounded by the $\omega^2=0$ contour at large $p_c^{\rm D}$ and at large $p_c^{\rm N}$, and the mass contours show that the maximum-mass configurations lie on this boundary, as expected. Superposed is the locus $\Delta R = 0$, which separates core from halo configurations and sweeps across the stable region as $\sigma/m_\chi$ increases. For $\sigma/m_\chi = 10\ {\rm cm^2/g}$, a larger portion of the stable, astrophysically relevant region lies on the halo side than for $0.1\ {\rm cm^2/g}$; among the stable configurations with $M \geq 1\,M_\odot$, the halo share grows from $30\%$ to $51\%$ and $57\%$ for $\sigma/m_\chi = 0.1$, $1$, and $10\ {\rm cm^2/g}$. This is the same trend quantified statistically in Sec.~\ref{subsec:RD_core_halo}, seen here at the level of the underlying configuration space rather than the posterior, and it confirms that the ordering found there is a property of the two-fluid structure equations and not an artifact of the sampling.
\section{Summary and Conclusions}
\label{sec:conclusion}
We have performed a Bayesian inference on two-fluid, dark-matter-admixed neutron stars in which the dark sector is an asymmetric, self-interacting fermion whose interaction strength is fixed by the elastic scattering cross section per unit mass measured on cluster and dwarf-galaxy scales. By repeating the full inference at $\sigma/m_\chi = 0.1$, $1$, and $10\ {\rm cm^2/g}$, spanning the astrophysically allowed range, against $\chi$EFT, perturbative QCD, the mass of PSR J0348+0432, five NICER pulsars, and GW170817, we have asked what this external observational input implies for neutron star structure. Our main conclusions are as follows.

(i) The nuclear matter parameters, the normal matter EOS, and the global observables $M_{\rm max}$, $R_{\rm max}$, $R_{1.4}$, and $\Lambda_{1.4}$ are statistically indistinguishable across two decades in $\sigma/m_\chi$, with $M_{\rm max}=2.07_{-0.05}^{+0.07}\,M_\odot$, $R_{1.4}\simeq12.1$~km, and $\Lambda_{1.4}\simeq385$ in every case. The medians agree to within $0.1$~km in $R_{1.4}$ and $9$ units in $\Lambda_{1.4}$, and the Bayesian evidence does not discriminate between the three cross sections ($\ln B = 0.30\pm0.05$). Sampling $\sigma/m_\chi$ as a free parameter gives the same results and shows that the present data are nearly uninformative on the cross section (Appendix~\ref{app:sigma_free}).

(ii) This insensitivity is not an accident of the parametrization. The self-interaction energy enters the dark EOS only as $(\sigma/m_\chi)^{1/2}$, and its effect on the star is further diluted by the small dark matter fraction. At fixed $f_\chi$, two decades in $\sigma/m_\chi$ shift the posterior baryonic radius by less than $0.01$~km at $1.4\,M_\odot$ and by at most about $0.1$~km near the maximum mass, far below the precision of present radius measurements.

(iii) The data prefer a small but non-zero dark matter fraction, $f_\chi = 0.05-0.09$, driven by the pulsars with the smallest radii and by GW170817, and bound it from above at $90\%$ credibility by $f_\chi < 0.13$, $0.16$ , and $0.19$ (90\% upper limits). The bound comes from the heavy pulsars through the maximum mass, and it loosens with the cross section; marginalizing over the cross section gives $f_\chi < 0.17$ (Appendix~\ref{app:sigma_free}).

(iv) The cross section is recorded in the dark sector and in the internal geometry. The posterior support for sub-GeV dark matter falls from $32\%$ to $21\%$ and $7\%$ as the cross section increases; the median dark core grows, its surface moving from $6.3$ to $3.1$~km inside the baryonic one at $1.4\,M_\odot$. The halo fraction rises from $8\%$ to $19\%$ at $1.4\,M_\odot$ and from $3\%$ to $10\%$ at $2\,M_\odot$ with the halo suppression mass rises from $1.14$ to $2.10\,M_\odot$ and the core-to-halo transition mass rises from $1.42$ to $1.58\,M_\odot$.

(v) Compared with a direct prior-predictive calculation, the astrophysical data reduce the odds of a halo configuration by a factor of two to three at $1.4\,M_\odot$ ($\mathcal{B}_{\rm halo} = 0.38$--$0.47$) and by a factor of nine to sixteen at $2\,M_\odot$ ($\mathcal{B}_{\rm halo} = 0.06$--$0.11$). Halo configurations that survive in the posterior fit the data as well as cores do.

(vi) The self-interaction is not intrinsically weak. In prior draws integrated at both extreme cross sections, so that only the cross section differs, it enlarges $R_{1.4}$ by up to $0.7$~km in the median, and by more than a kilometer for large dark fractions. These effects are concentrated at $m_\chi \lesssim 1.5$~GeV, precisely the region whose posterior weight the data remove as the cross section increases.

Several extensions are possible, such as the I--Love--Q/C universal relations of two-component stars and the non-radial oscillation spectrum, in which a two-fluid configuration supports mode families with no single-fluid counterpart. Finally, a Yukawa-mediated dark sector, in which the cross section acquires a velocity dependence, would break the degeneracy between $\sigma$ and $m_\chi$. We intend to pursue these directions in future work.
\section*{Acknowledgments}
HCD is deeply grateful to J\"urgen Schaffner-Bielich for his continuous guidance, insightful suggestions, and the detailed discussions throughout the course of this project, which significantly shaped this work. HCD also thanks Hans-Josef~Schulze for helpful discussions regarding the cross section of SIDM. HCD acknowledges support from the Alexander von Humboldt Foundation.
\section*{Data Availability}
The data that support the findings of this article are not publicly available upon publication because it is not technically feasible to make them available within the terms of this research project.
\appendix
\setcounter{figure}{0}
\setcounter{table}{0}
\renewcommand{\thefigure}{A\arabic{figure}}
\renewcommand{\thetable}{A\arabic{table}}
\renewcommand{\theHfigure}{A\arabic{figure}}
\renewcommand{\theHtable}{A\arabic{table}}
\section{Cross section as a free parameter}
\label{app:sigma_free}
In the main text, the inference is repeated at three fixed values of $\sigma/m_\chi$. Here we treat the cross section as an additional free parameter, with a log-uniform prior on $[0.1, 10]$~cm$^2$/g, and keep all other priors (Table~\ref{tab:prior_table}), the likelihood (Eq.~(\ref{eq:likelihood})), and the sampler settings unchanged. For each draw, the interaction parameter $y$ then follows from Eq.~(\ref{eq:ymu}) with the sampled $\sigma/m_\chi$ and $m_\chi$. This run answers two questions that the fixed-$\sigma$ runs leave open; whether the data constrain the cross section itself, and what the dark sector bounds are once the cross section is marginalized over.
\begin{figure}[h]
\centering
\includegraphics[width=\columnwidth]{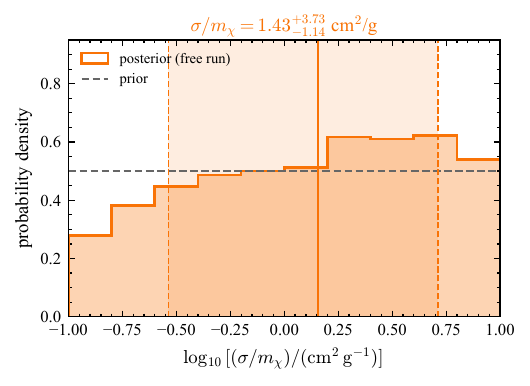}
\caption{Marginalized posterior of the dark matter self-interaction cross section $\sigma/m_\chi$ when it is sampled as a free parameter (orange), compared with its log-uniform prior on $[0.1, 10]$~cm$^2$/g (grey dashed). The solid vertical line marks the median and the shaded band, bounded by the orange dashed lines, the $68\%$ CI, whose values are quoted above the panel.}
\label{fig:sigma_free}
\end{figure}

Figure~\ref{fig:sigma_free} shows the posterior of $\sigma/m_\chi$. It departs only moderately from the prior, with a preference for larger cross sections that is concentrated at the low end: the lowest half-decade, $0.1$--$0.32$~cm$^2$/g, carries $17.5\%$ of the posterior compared with $25\%$ of the prior, while the other three carry $24.4$--$29.5\%$. We find $\sigma/m_\chi = 1.43^{+3.73}_{-1.14}$~cm$^2$/g ($68\%$ CI), the $90\%$ CI is $0.21$--$6.52$~cm$^2$/g, and $58\%$ of the posterior lies above $1$~cm$^2$/g. This preference is in the same direction as the small evidence ratio between the fixed-$\sigma$ runs, $\ln\mathcal{Z}(10) - \ln\mathcal{Z}(0.1) = 0.30$, but it is weak: the present data are essentially uninformative on the cross section, and its posterior is dominated by the prior range, which we take from cluster and dwarf-galaxy observations.

\begin{table}
\centering
\caption{Median values with $68\%$ CIs from the run with $\sigma/m_\chi$ free, compared with the fixed run at $\sigma/m_\chi = 1$ cm$^2$/g. The upper limit on $f_\chi$ is the one-sided $90\%$ limit. Radii are those of the baryonic surface $R_{\rm N}$.}
\label{tab:sigma_free}
{\renewcommand{\arraystretch}{1.5}
\begin{tabular}{lcc}
\hline\hline
Quantity & $\sigma/m_\chi$ free & $\sigma/m_\chi = 1$ cm$^2$/g \\
\hline
$\sigma/m_\chi$ (cm$^2$/g) & $1.43^{+3.73}_{-1.14}$ & $1$ (fixed) \\
$n_{\rm sat}$ (fm$^{-3}$)  & $0.160^{+0.007}_{-0.009}$ & $0.161^{+0.006}_{-0.009}$ \\
$E_{\rm sat}$ (MeV)        & $-16.05^{+0.34}_{-0.30}$ & $-16.05^{+0.35}_{-0.29}$ \\
$K_{\rm sat}$ (MeV)        & $244.00^{+34.63}_{-27.75}$ & $240.22^{+35.30}_{-27.10}$ \\
$J_{\rm sat}$ (MeV)        & $31.89^{+3.31}_{-3.31}$ & $31.29^{+3.76}_{-3.35}$ \\
$L_{\rm sat}$ (MeV)        & $50.49^{+9.05}_{-6.72}$ & $50.39^{+9.61}_{-6.52}$ \\
$M^*_{\rm sat}$            & $0.64\pm0.05$ & $0.64\pm0.05$ \\
\hline
$m_\chi$ (GeV)             & $1.71^{+1.26}_{-0.74}$ & $1.53^{+1.24}_{-0.63}$ \\
$f_\chi$                   & $0.08^{+0.07}_{-0.04}$ & $0.07^{+0.07}_{-0.04}$ \\
$f_\chi$ ($90\%$ upper limit) & $0.17$ & $0.16$ \\
$P(m_\chi < 1\ {\rm GeV})$ & $17.6\%$ & $21.3\%$ \\
\hline
$M_{\rm max}$ ($M_\odot$)  & $2.07^{+0.06}_{-0.05}$ & $2.07^{+0.07}_{-0.05}$ \\
$R_{\rm max}$ (km)         & $11.09^{+0.32}_{-0.31}$ & $11.07^{+0.34}_{-0.32}$ \\
$R_{1.4}$ (km)             & $12.13^{+0.30}_{-0.27}$ & $12.13^{+0.31}_{-0.29}$ \\
$\Lambda_{1.4}$            & $386^{+82}_{-62}$ & $385^{+84}_{-64}$ \\
\hline\hline
\end{tabular}}
\end{table}

Table~\ref{tab:sigma_free} compares the free run with the fixed run at $\sigma/m_\chi = 1$ cm$^2$/g. The nuclear matter parameters and the global observables agree with those of every fixed-$\sigma$ run to well within their CIs, which confirms the discussion of Sec.~\ref{subsec:RD_NMPs}; i.e., the nuclear sector is insensitive to the cross section whether it is fixed or marginalized over. The dark sector parameters, which do depend on $\sigma/m_\chi$ (Sec.~\ref{subsec:RD_DMPs}), lie between the values of the fixed runs. Marginalizing over the cross section, the data give $f_\chi = 0.08^{+0.07}_{-0.04}$ with a $90\%$ upper limit $f_\chi < 0.17$, and a posterior probability of $17.6\%$ for sub-GeV dark matter, compared with $f_\chi < 0.13-0.19$ and $7.4-31.6\%$ across the fixed runs. These marginalized values are the appropriate summary of the dark sector when the cross section is regarded as unknown within its astrophysically allowed range.

\bibliography{twofluid}
\bibliographystyle{apsrev4-2}
\end{document}